\documentstyle[aps]{revtex}

\begin{document}

\author{Alkan Kabak\c c\i o\u{g}lu$^{1,2}$and A. Nihat Berker$^{1,2,3}$}
\address{$^1$Department of Physics, Massachusetts Institute of Technology\\
Cambridge, Massachusetts 02139, U.S.A.\\
$^2$Feza G\"ursey Research Center for Basic Sciences\\
\c Cengelk\"oy, Istanbul 81220, TURKEY\\
$^3$Department of Physics, Istanbul Technical University\\
Maslak, Istanbul 80626, TURKEY}
\title{Strongly Asymmetric Tricriticality of Quenched Random-Field Systems }
\maketitle

\begin{abstract}
In view of the recently seen dramatic effect of quenched random bonds on
tricritical systems, we have conducted a renormalization-group study on the
effect of quenched random fields on the tricritical phase diagram of the
spin-1 Ising model in $d=3$. We find that random fields convert first-order
phase transitions into second-order, in fact more effectively than random
bonds. The coexistence region is extremely flat, attesting to an unusually
small tricritical exponent $\beta _u$; moreover, an extreme asymmetry of the
phase diagram is very striking. To accomodate this asymmetry, the
second-order boundary exhibits reentrance.
\end{abstract}
\pacs{}

\narrowtext

Tricritical phase diagrams of three-dimensional $(d=3)$ systems are strongly
affected by quenched bond randomness: The first-order phase transitions are
replaced, gradually as randomness is increased, by second-order phase
transitions. The intervening random-bond tricritical point moves
towards, and eventually reaches, zero-temperature, as the amount of
randomness is increased. This
behavior is an illustration of the general prediction that first-order phase
transitions are converted to second-order by bond randomness
\cite{Wehr,Hui,ANBApplied,ANBPhysica,FBTurk,Chen92,Chen95}, in a
thresholded manner in
$d=3$.  The randomness threshold increases from zero at the non-random
tricritical point. The random-bond tricritical point maps, under
renormalization-group transformations, onto a doubly unstable fixed
distribution at
strong coupling. Random-bond tricritical points exhibit a remarkably small
value for
the tricritical exponent $\beta _u=0.02$, reflected in the near-flat top of the
coexistence region. In the conversion of the first-order phase transition to
second-order, traced by the random-bond tricritical point, a strong
violation of the
empirical universality principle occurs, via a renormalization-group fixed-point
mechanism.   Thus, detailed information now exists on the effect of
quenched {\it
bond} randomness on tricritical points, revealing several qualitatively
distinctive features.\cite{FBTurk,FB}

No such information has existed on the effect of
quenched {\it field} randomness on tricritical points. This topic is of
interest also because renormalization-group studies have shown that quenched
field randomness induces, under scale change, quenched bond randomness, as
the presence of quenched field randomness \mbox{continues.}\cite{McB}
Accordingly,
we have conducted a global renormalization-group study of a tricritical
system in $d=3$ under quenched random fields. The results, presented below,
show that these systems have their own distinctive behavior which is
qualitatively different from that of non-random or random-bond tricritical
systems. The latter distinction yields a microscopic physical intuition on
the different effects of the two types of quenched randomness.

We have studied the Blume-Emery-Griffiths (i.e., spin-1 Ising) model under
quenched field randomness. The Hamiltonian is
\begin{eqnarray}
-\beta {\cal H}=
&&\sum_{<ij>}
        \left[ \rule{0in}{.15in}
        Js_is_j+Ks_i^2s_j^2-\Delta
        \left(
        s_i^2+s_j^2
        \right)
        \right. \nonumber \\
        & + & \left. H_{ij} \left( s_i+s_j \right) + H_{ij}^{\dagger}\left(
s_i-s_j \right)
        \right] \,\, ,
\end{eqnarray}
where $s_i=\pm 1,0$ at each site $i$ of a simple cubic $(d=3)$ lattice and
$<ij>$ indicates summation over all nearest-neighbor pairs of sites. The
quenched random fields $H_{ij},H_{ij}^\dagger$ are taken from a distribution
\widetext
\FL
\begin{eqnarray}
P(H,H^\dagger) & = & \frac{1}{4}
\left[ \rule{0in}{.15in}
\delta \left( H + \sigma _H \right) \delta \left( H^\dagger + \sigma _{H}
\right)
\right. \nonumber \\
& + & \left. \delta (H+\sigma _H)\delta \left( H^\dagger-\sigma _{H}
\right) \right.
 + \delta (H-\sigma
_H)\delta \left( H^\dagger+\sigma _{H} \right)
+ \left. \delta \left( H-\sigma _H \right) \delta \left
(H^\dagger-\sigma_{H} \right)
\rule{0in}{.15in} \right]
\,\,\, .
\end{eqnarray}
\narrowtext
\noindent
All other interactions in the initial Hamiltonian (1) are non-random. Under
renormalization-group transformations, the Hamiltonian (1) maps onto a
random-field random-bond Hamiltonian,

\widetext
\FL
\begin{eqnarray}
-\beta {\cal H} & = & \sum_{<ij>}
\left[
 J_{ij}s_is_j  +   K_{ij}s_i^2s_j^2-\Delta _{ij} ( s_i^2+s_j^2 )
-\Delta_{ij}^\dagger ( s_i^2-s_j^2 )
\right.  \nonumber \\
& + & \left.
L_{ij} \left( s_i^2s_j+s_is_j^2 \right) + L_{ij}^{\dagger}
\left( s_i^2s_j-s_is_j^2 \right)  + H_{ij} \left( s_i+s_j \right) +
H_{ij}^{\dagger}(s_i-s_j)
\right]
 \,\,\, ,
\end{eqnarray}
\narrowtext
\noindent
where all interactions are quenched random, with a distribution function
$
P\left( J_{ij},K_{ij},\Delta _{ij},\Delta _{ij}^{\dagger},L_{ij},L_{ij}
^{\dagger},H_{ij},H_{ij}^{\dagger} \right)
$
determined by the renormalization-group
transformation. Specifically, the first four arguments here reflect the
rescaling-induced bond randomness of the random-field system.

The renormalization-group transformation is contained in the mapping between
the quenched probability distributions of the starting and rescaled systems,
\widetext
\FL
\begin{equation}
P^{\prime } \left( \overrightarrow{K}_{i^{\prime }j^{\prime }}^{\prime} \right)
= \int
\left[ \prod^{<i^\prime j^\prime >}_{<ij>}
d
\overrightarrow{K}_{ij}P \left( \overrightarrow{K}_{ij} \right) \right]
\delta \left(
\overrightarrow{K}_{i^{\prime }j^{\prime }}^{\prime }-\overrightarrow{R} \left(
\left\{
\overrightarrow{K}_{ij} \right\} \right) \right) \,\,\, ,
\end{equation}
\narrowtext
\noindent
where $\overrightarrow{K}_{ij}\equiv (J_{ij},K_{ij},\Delta _{ij},\Delta
_{ij}^{\dagger },L_{ij},L_{ij}^{\dagger },H_{ij},H_{ij}^{\dagger }) $ are the
interactions at locality $<ij>$, the primes refer to the renormalized
system, the product is over all unrenormalized localities $<ij>$ whose
interactions
$
\{ \overrightarrow{K}_{ij} \}
$
influence the renormalized
interaction
$\overrightarrow{K}_{i^{\prime }j^{\prime }}^{\prime }
$,
and
$
\overrightarrow{R} ( \{ \overrightarrow{K}_{ij} \} )
$
is a local recursion
relation that embodies the latter dependence. Simply said, Eq.(4) sums over
the joint probabilities of the values of neighboring unrenormalized
interactions that conspire to yield a given value of the renormalized
interaction. The phenomena characteristic to quenched randomness should
derive from the probability convolution shown in Eq.(4), rather than the
precise form of the recursion $\overrightarrow{R}$ that should be a smooth
local function. In this work, we use the Migdal-Kadanoff recursion relation,
given for this system in \cite{LB}. The convolution is effected by
representing $P ( \overrightarrow{K}_{ij} ) $ in terms of bins, the degree of
detail (i.e., the number of bins) reflecting the level of approximation. In
this work, we have used 531,441 bins, corresponding to renormalization-group
flows in a 4,782,969-dimensional space. The application of Eq.(4) via the
binning procedure has been described elsewhere \cite{FB,FBM}.

Our main result, the striking difference between the three types of $d=3$
Ising tricriticality, is evident in Fig.1, where the calculated
random-field, random-bond, and non-random phase diagrams are superimposed.
The same amount of quenched randomnes [$\sigma _H=0.2=\sigma _\Delta $, as
in Eq.(2)] is used, for relevance of comparison. It is seen that both bond
randomness  and field randomness convert first-order phase
transitions to second-order in a thresholded manner, but that field
randomness is more effective than bond randomness in this conversion. Both
types of random tricritical points occur at remarkably near-flat tops of
coexistence regions, reflecting the unusually small values of the exponent $
\beta _u,$ but the random-field phase diagram is most strikingly
asymmetrical. The tricritical point occurs at the high density $<s_i^2>\, =
0.835
$ (as opposed to $0.613$ and $0.665$ in the random-bond and non-random
systems, respectively), essentially all of the near-flat portion of the
coexistence top occurring on the dilute branch of the coexistence boundary.
To accomodate this asymmetry, the randomness-extended line of second-order
phase transitions has to curve over and exhibit reentrant behavior \cite
{Cladis} as a function of temperature.  This difference in behavior comes
from the fact that
random bonds destroy order-disorder coexistence without destroying order itself
\cite{ANBApplied,ANBPhysica}, whereas random fields destroy both
order-disorder coexistence, through
the rescale-generated bond randomness, and order per se.  The latter is
more effective near the
tricritical point, where considerable vacancy fluctuations occur within the
ordered phase.

Random-field and random-bond tricritical points renormalize onto obviously
different doubly unstable fixed distributions (the field variables $L$ and
$H$ remain at zero in the latter case).

It is seen in Fig.1 that the coexistence boundary of either type of random
system follows that of the non-random system, until the temperature-lowered
tricritical region sets in relatively abruptly. This is similar to the
magnetization of random-field systems following the non-random curve until
the critical region sets in quite abruptly.\cite{Machta,FBM}

On the high-temperature side of the tricritical point, it has been known
\cite{FB}
that the break in slope of the critical line, in the random-bond system, is
connected to the strong violation of the empirical universality principle,
segments on each side of this point having their own critical exponents,
respectively of the strong-coupling and non-strong-coupling type. No such
universality violation occurs along the second-order line of the
random-field system, the entire line having random-field critical exponents
that are governed by a strong-coupling fixed distribution, implying a
modified hyperscaling relation \cite{BM,BMc}.

The global phase diagrams of the random-field and random-bond systems are
given in Figs.2(a,b). It is seen that all first-order phase transitions are
completely converted to second-order, at a zero-temperature tricritical
point, for $\sigma _H\simeq 0.5$ and $\sigma _\Delta \simeq 0.7$,
respectively. Furthermore, the ordered phase is eliminated at
$\sigma_H \simeq 0.9$ in the random-field case, but persists for all for $\sigma
_\Delta $ in the random-bond case.

This research was supported the U.S. Department of Energy under Grant No.
DE-FG02-92ER45473 and by the Scientific and Technical Research Council of
Turkey (T\"UBITAK).  We gratefully acknowledge the hospitality of the Feza
G\"{u}rsey Research
Center for Basic Sciences and of the Istanbul Technical University.

\begin{center}
{\bf Figure Captions}
\end{center}
\bigskip
\bigskip
\noindent
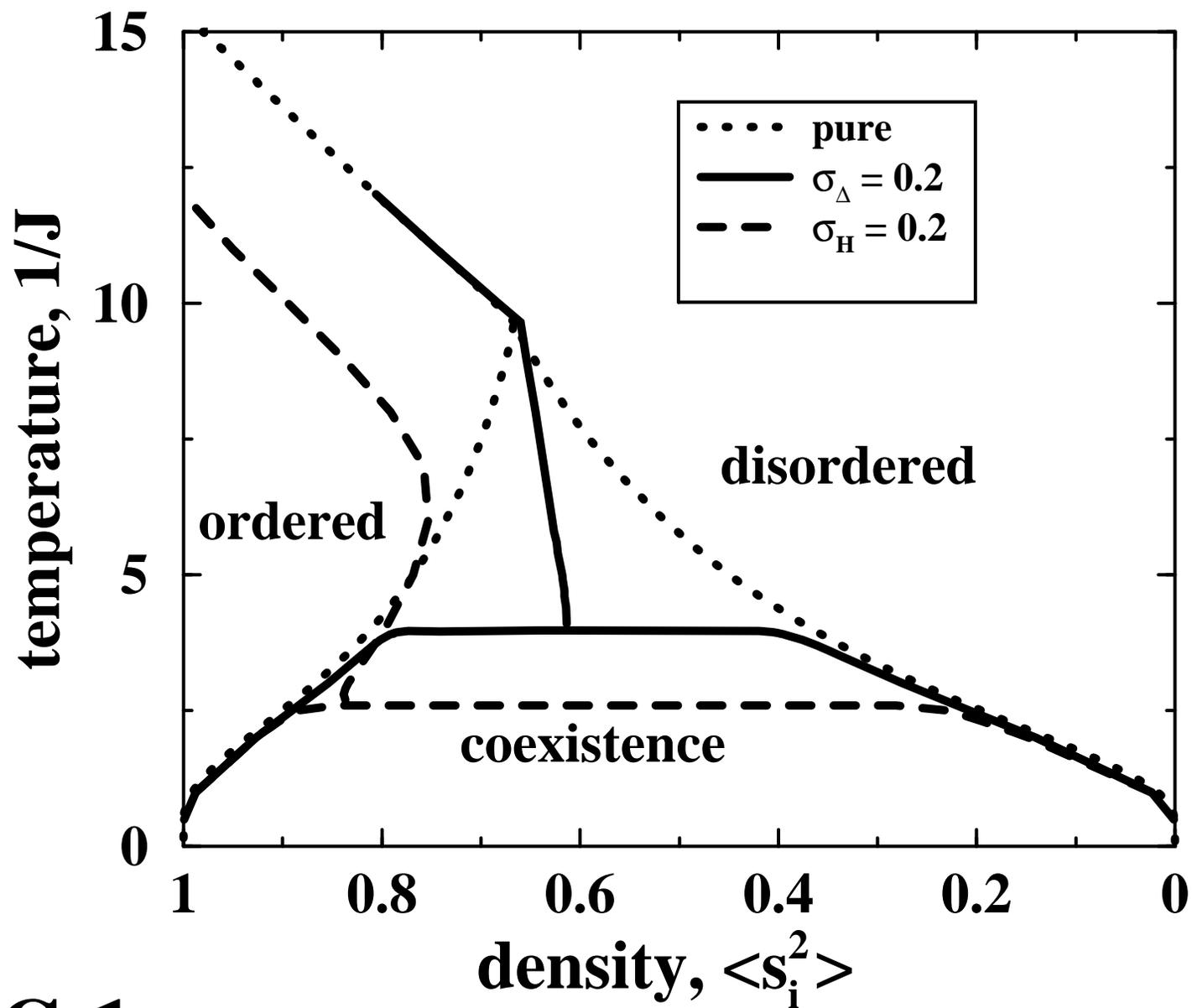
{\bf Fig. 1}: Calculated tricritical phase diagrams for non-random
(dotted), random-bond (full),
and random-field (dashed) $d = 3$ systems for $K = 0$.  In each phase
diagram, a line of
second-order phase transitions extending to high temperatures  meets, at a
tricritical point, a
coexistence region extending to low temperatures.  Note the near-flat top
of the coexistence regions
in both quenched random systems, and the extreme asymmetry of the
random-field system.  Thus, field
randomness is more effecive than bond randomness in converting first-order
transitions into
second-order (i.e., the tricritical point is at lower temperature).

\bigskip
\bigskip
\noindent
{\bf Figs 2}:  Calculated $d = 3$ global phase diagrams for $K = 0$:
(a)~Random-bond systems
exhibit two universality classes of second-order phase transitions (thin
and thick full lines) and
first-order phase transitions (circles). (b)~Random-field systems exhibit
second-order (full lines)
and first-order (circles) phase transitions.  In both types of quenched
randomness, the first-order
transitions cede under increased randomness.  The line of tricritical
points (dashed) reaches zero
temperature, as all transitions become second-order. In the random-field
case, the ordered phase
disappears under further randomness, whereas in the random-bond case, the
ordered phase (and the
strong violation of universality) persists for all randomness.

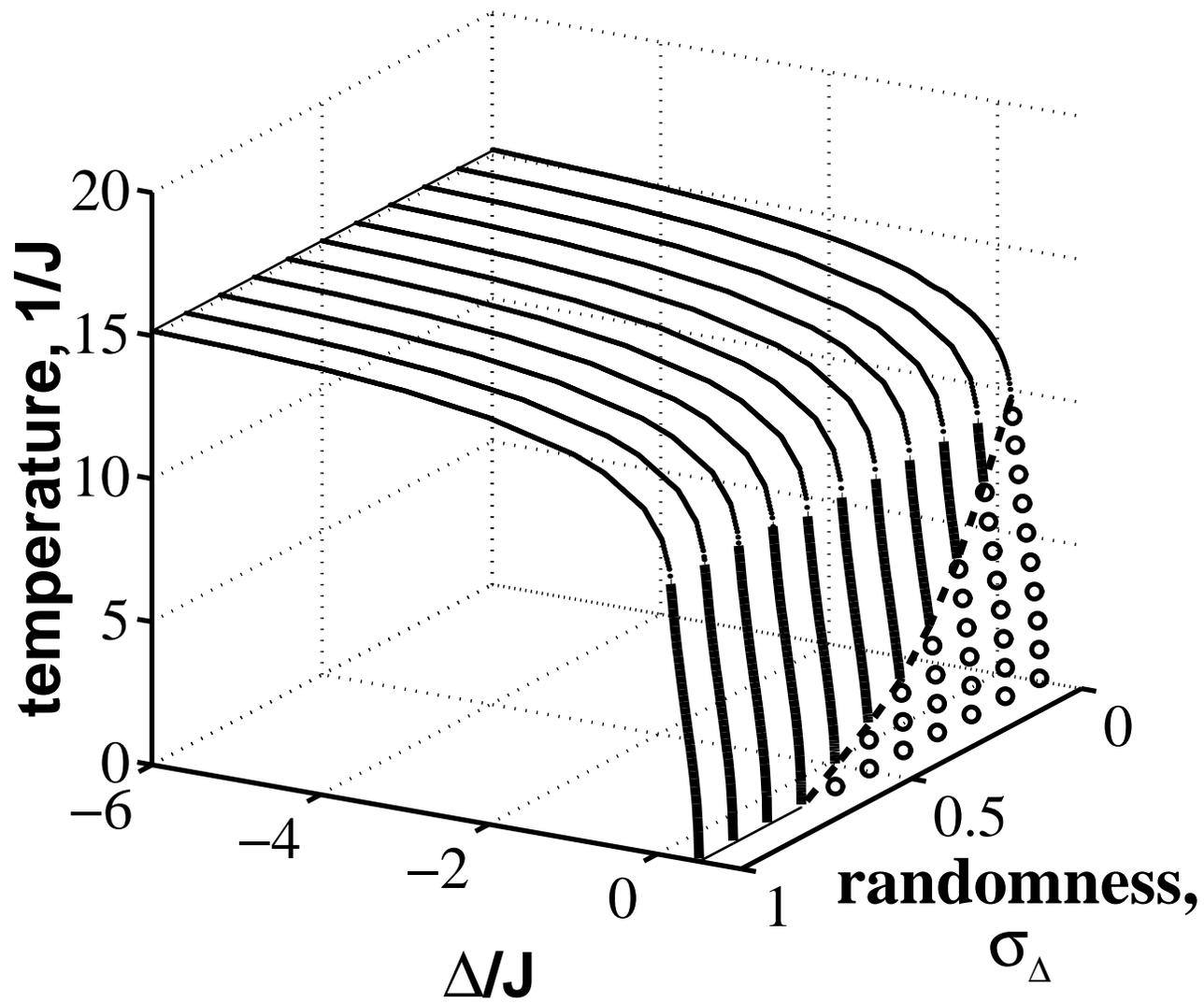
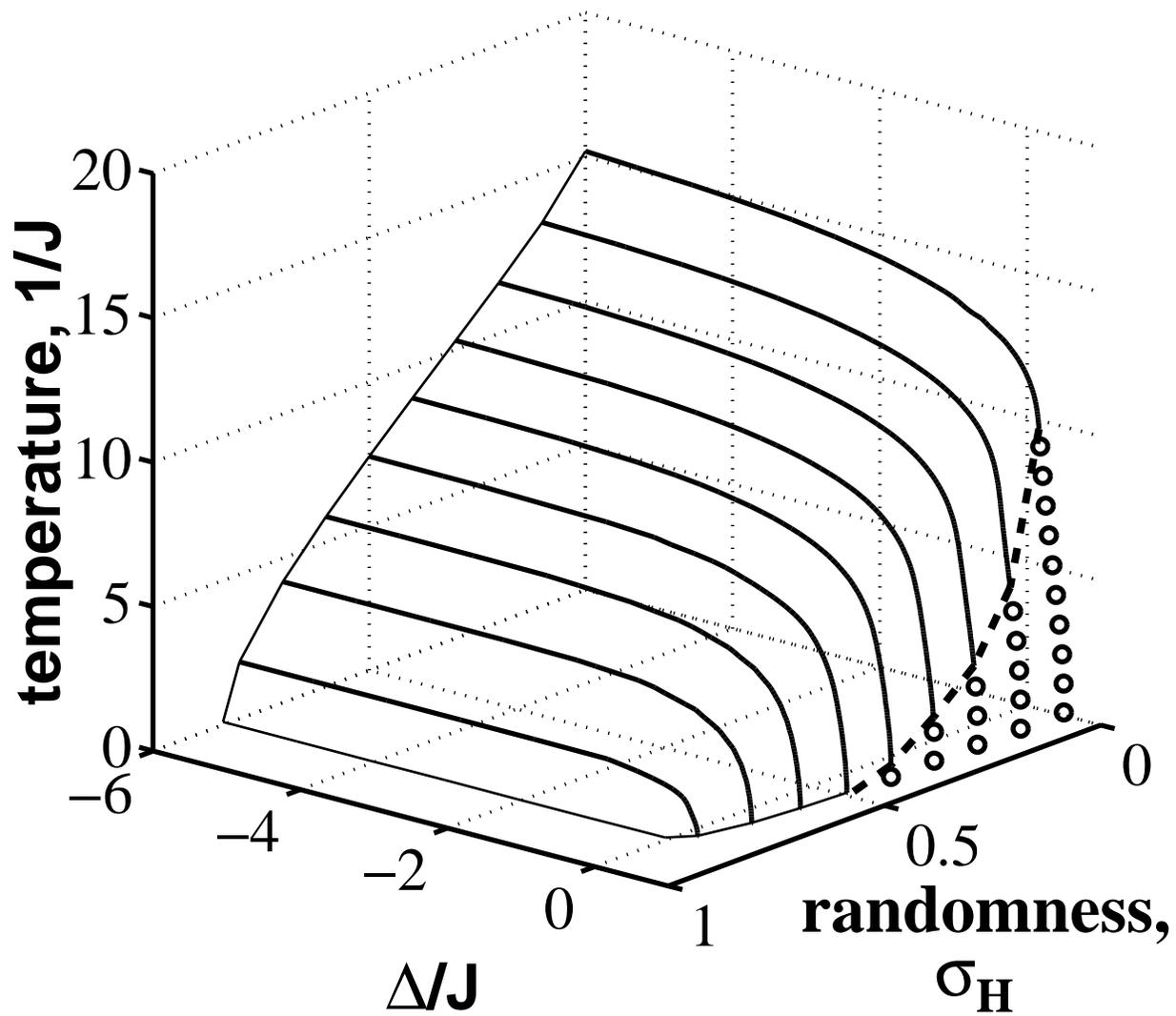
\end{document}